\documentstyle[amssymb,aps]{revtex}
\begin{document}
\title{Entanglement bases and general structures of orthogonal complete bases}
\author{Zai-Zhe Zhong}
\address{Department of Physics, Liaoning Normal University, Dalian 116029, Liaoning,\\
China. E-mail: zhongzaizheh@hotmail.com}
\maketitle

\begin{abstract}
In quantum mechanics and quantum information, to establish the orthogonal
bases is a useful means. The existence of unextendible product bases impels
us to study the `entanglement bases' problems. In this paper, the concepts
of entanglement bases and exact-entanglement bases are defined, and a
theorem about exact-entanglement bases is given. We discuss the general
structures of the orthogonal complete bases. Two examples of applications
are given. At last, we discuss the problem of transformation of the general
structure forms.

PACC numbers: 03.67.Mn, 03.65.Ud, 03.67.Hk.
\end{abstract}

In quantum mechanics and quantum information, the concept of bases of a
Hilbert space is of paramount importance. A product basis (PB) of a Hilbert
space of multipartite quantum states is a set of orthogonal product
pure-states. In common cases, we always use the standard bases, which are
the simplest PBs. However, especially in quantum information theory, one
need to use the bases consisting of some entangled pure-states. For
instance, in the study of teleportation the Bell's bases, which are
maximally entangled states, are very important. Thus, it is necessary that
in quantum information we consider various possible orthogonal bases. In
principle, a general orthogonal complete basis consists of some product
pure-states and some entangled pure-states. Related this, recently a problem
discussed in depth is the so-called `unextendible product basis (UPB)'[1],
which must relate to the bases consisting of some entangled pure-states.
There are many works about UPB, mainly see Bennett, DiVincenzo, Terhal et
al.[1-5$],$ in addition, [6-12], these works show that the UPBs have many
quite special properties$.$

An UPB of a Hilbert space $H$ of multipartite quantum states is a PB $S$, $S$
spans a subspace $H_S$ in $H$, and the complementary subspace $H{\cal -}H_S$
contains no product state. The existence of UPBs means that the structures
of general orthogonal complete bases are more complex. Since either a
pure-state is a product state, or an entangled state (it may be partial
separable), to discuss the structures of general orthogonal complete bases
of $H$ is just to discuss how the product pure-states and entangled
pure-states assort in the bases. In this paper, first we define the concepts
of entanglement basis and exact-entanglement basis, and give a theorem about
the exact-entanglement bases. Two examples of applications are given. We
discuss the problem of the general structures of the orthogonal complete
bases. At last, we consider the transformation problem.

Generally, for a given pure-state we can more easily decide whether it is
separable (i.e. a product state), or is entangled. For instance, according
to the entanglement of formation[$13],$ only the rank of the coefficient
matrix of a bipartite qubit pure-states can decide whether this state is
separable, or not. As for a given multipartite qubit pure-state, we can use
other criteria, e.g. the criteria given by D\"{u}r[14$]$ and Kauffman et al.[%
$15],$ etc.. Since the purpose in this paper is not to study the problem of
criteria of separability, for the sake of simplicity, in the following we
assume that for any given multipartite pure-state we always can decide
whether it is a product state, or is an entangled state.

Consider a multipartite quantum system $H{\cal =\otimes }_{i=1}^MH_i$\ with $%
M$\ parties of respective dimension $d_i$, the total dimensionality of $H$
is $N=\prod\limits_{i=1}^Md_i.$ If an orthogonal complete basis $B$ =$%
\left\{ \mid \omega _0>,\mid \omega _1>,\cdots ,\mid \omega _{N-1}>\right\} $
of $H$ is given, then we always have 
\begin{equation}
\sum\limits_{s=0}^{N-1}\mid \omega _s><\omega _s\mid =I
\end{equation}
where $I$ is the $N\times N$ unit matrix. $B$ can be split into two parts as 
$S{\cal =}\left\{ \mid \psi _0>,\cdots ,\mid \psi _{m-1}>\right\} $ and $T=$%
{\it $\left\{ \mid \varphi _0>,\cdots ,\mid \varphi _{n-1}>\right\} $, }i.e. 
$\mid \psi _i>=\mid \omega _i>,\mid \varphi _j>=\mid \omega _{j+m-1}>\left(
m+n=N\right) $, then $B=S\cup T=\left\{ \mid \psi _0>,\cdots ,\mid \psi
_{m-1}>,\mid \varphi _0>,\cdots ,\mid \varphi _{n-1}>\right\} .$ Let $H_S$
and $H_T$, respectively, be the subspaces spanned by $S{\cal \ }$and $T.$
For any orthogonal basis $S^{\prime }$ of $H_S,$ there must be a $m\times m$
unitary matrix $U\left( m\right) $ such that $S^{\prime }=S_{U\left(
m\right) }=\left\{ \psi _0^{\prime },\cdots ,\psi _{m-1}^{\prime }\right\} ,$
where $\mid \psi _i^{\prime }>=\sum\limits_{s=0}^{m-1}\left[ U\left(
m\right) \right] _{is}\mid \psi _s>.$ Similarly, any orthogonal basis $%
T^{\prime }$ of $H_T$ can be written as $T^{\prime }=T_{U\left( n\right)
}=\left\{ \varphi _0^{\prime },\cdots ,\varphi _{n-1}^{\prime }\right\} ,$
where $\mid \varphi _j^{\prime }>=\sum\limits_{s=0}^{n-1}\left[ U\left(
n\right) \right] _{js}\mid \varphi _s>.$

{\bf Definition.} {\it Consider a multipartite quantum system }$H{\cal %
=\otimes }_{i=1}^MH_i${\it \ with }$M${\it \ parties of respective dimension 
}$d_i$, {\it the total dimensionality of} $H$ {\it is }$N=\prod%
\limits_{i=1}^Md_i.${\it \ An entanglement basis (EB) }$T=${\it $\left\{
\mid \varphi _0>,\cdots ,\mid \varphi _{n-1}>\right\} $ is a set of n
entangled pure-states, }$\mid \varphi _j>\left( j=0,\cdots ,n-1\right) ,$ 
{\it such that an arbitrary linear combination of them steel is an entangled
pure-state. The subspace }$H_T${\it \ spanned by an EB }$T${\it \ ( }$H_T$%
{\it \ does not contain any disentangled pure-states) is called an
entanglement space (ES). An EB }$T${\it \ is called exact-entanglement basis
(EEB) if there is a UPB }$S=\left\{ \mid \psi _0>,\cdots ,\mid \psi
_{m-1}>\right\} ${\it \ \ containing }$m=N-n${\it \ product states such that 
}$B=S\cup T=\left\{ \mid \psi _0>,\cdots ,\mid \psi _{m-1}>,\mid \varphi
_0>,\cdots ,\mid \varphi _{n-1}>\right\} ${\it \ forms an orthogonal
complete basis of }$H${\it . In this case the subspace }$H_T${\it \ is
called an exact-entanglement space (EES), in which all states and the UPB }$%
S ${\it \ are orthogonal each other. And we call }$B${\it \ a complete basis
with an unextendible product basis (CBUPB).}

For an EES $H_{ES}${\it \ , }the corresponding UPB $S${\it \ }is in $H${\it $%
_{ES}^{\bot },$ }where{\it \ }$H_{ES}^{\bot }${\it \ }is{\it \ }the
orthogonal complementary subspace of $H_{ES}${\it $.$ }About this, recently
Bravy[11] proves a result in the case of tripartite qubit states (the Lemma
3 in [$11$]). However, it is easily seen that the Lemma 3 and its proof in
[11], in fact, can be directly generalized to higher dimensional cases, thus
we have the following useful lemma

{\bf Lemma 1.}{\it \ For an EES }$H_{ES}${\it $,$} {\it the corresponding
UPB in }$H_{ES}^{\bot }${\it \ is unique.}

As in the above mention, it is necessary that in quantum information we
should consider the general orthogonal complete bases. In the first place,
we consider the problem what are the conditions of an EB to be an EEB.
Another problem is how to describe the general form of orthogonal complete
bases. The following theorem can answer the above problems, it$,$in fact, is
a criterion of the existence of EEBs, and it can be used to describe the
general structures of orthogonal complete bases.

{\bf Theorem 1}. {\it The sufficient and necessary conditions for an ES }$%
H_{ES}${\it \ to be an EES are that }

$\left( A\right) ${\it \ The dimensionality }$n$ {\it of }$H_{ES}$ {\it %
satisfies }$n\leq N-\sum\limits_{i=1}^M\left( d_i-1\right) -1.\;$

$\left( B\right) $ {\it For any orthogonal basis }$T=\left\{ \mid \varphi
_0>,\cdots ,\mid \varphi _{N-n-1}>\right\} $ {\it of }$H_{ES}$ {\it , the
uniform mixture }$\stackrel{\_}{\rho }=\frac 1{N-n}\left\{
I-\sum\limits_{j=0}^{n-1}\mid \varphi _j><\varphi _j\mid \right\} ${\it \ is
separable, where }$I$ {\it is the }$N\times N${\it \ unit matrix.}

{\bf Proof. Necessity. }Suppose that $H_{ES}${\it \ is }a $n$-dimensional
EES and $T=\left\{ \mid \varphi _0>,\cdots ,\mid \varphi _{n-1}>\right\} $
is an arbitrary orthogonal basis in $H_{ES},$ then there is an UPB $S$=$%
\left\{ \mid \psi _0>,\cdots ,\mid \psi _{m-1}>\right\} ,$ every $\mid \psi
_i>$ $(i=0,\cdots ,m-1,\;m+n=N)$ is a product pure-states, such that $S\cup
T $ forms a CBUPB of $H.$

$\left( A\right) $ According to [1], the number $m$ of states in an UPB must
satisfy $m\geq $ $\sum\limits_{i=1}^M\left( d_i-1\right) +1,$ then $%
n=N-m\leq N-\sum\limits_{i=1}^M\left( d_i-1\right) -1.$

$\left( B\right) $ According to Eq.(1), we have 
\begin{equation}
\sum_{i=0}^{m-1}\mid \psi _i><\psi _i\mid +\sum_{j=0}^{n-1}\mid \varphi
_j><\varphi _j\mid =I
\end{equation}
Here $\mid \psi _0>,\cdots ,\mid \psi _{m-1}>$ all are products of
pure-states, according to the definition of separability of the multipartite
systems, the uniform mixture 
\begin{equation}
\stackrel{\_}{\rho }=\frac 1{N-n}\left\{ I-\sum\limits_{j=0}^{n-1}\mid
\varphi _j><\varphi _j\mid \right\} =\sum_{i=0}^{m-1}\frac 1m\mid \psi
_i><\psi _i\mid
\end{equation}
is separable.

{\bf Sufficiency}. Suppose that for a subspace $H_{ES}$ the conditions (A)
and (B) both hold. Now, in this $H_{ES}$ we take arbitrarily an orthogonal
basis $T=\left\{ \mid \varphi _0>,\cdots ,\mid \varphi _{n-1}>\right\}
(n\leq N-\sum\limits_{i=1}^M\left( d_i-1\right) -1),$ then according to (B),
the density matrix{\it \ }$\stackrel{\_}{\rho }=\frac 1{N-n}\left\{
I-\sum\limits_{j=0}^{n-1}\mid \varphi _j><\varphi _j\mid \right\} $\ is
separable, i.e. there must be a decomposition of {\it \ }$\stackrel{\_}{\rho 
}$ as $\frac 1{N-n}\left\{ I-\sum\limits_{j=0}^{n-1}\mid \varphi _j><\varphi
_j\mid \right\} =\stackrel{\_}{\rho }=\sum_sp_s\rho _s$, where $0<p_s\leq
1,\sum\limits_sp_s=1,$ every $\rho _s$=$\mid \Psi _s><\Psi _s\mid $ is a
product state$.$ Then we have 
\begin{equation}
\;\frac 1n\sum\limits_{j=0}^{n-1}\mid \varphi _j><\varphi _j\mid =\frac 1n%
\left\{ I-\left( N-n\right) \sum_sp_s\rho _s\right\}
\end{equation}
Right multiplied two sides of Eq.($4$) by $\mid \Psi _t>,$ we obtain 
\begin{equation}
\sum_{j=0}^{n-1}\lambda _{j,t}\mid \varphi _j>=\sum_s\mu _{s,t}\mid \Psi _s>,%
\text{ }
\end{equation}
where $\lambda _{j,t}=\frac 1n<\varphi _j\mid \Psi _t>$ and $\mu _{s,t}=%
\frac 1n\left( \delta _{st}-\left( N-n\right) p_s\right) <\Psi _s\mid \Psi
_t>.$ But according to the definition of EB and the above supposition,
generally Eq.($5$) is impossible (the left side is entangled, but the right
side is separable), only possibility is that two sides of Eq.($5$) both
vanish. From $\lambda _{j,t}\equiv 0,$ then $<\varphi _j\mid \Psi
_t>=n\lambda _{j,t}\equiv 0,$ i.e. all $\mid \Psi _t>$ are orthogonal to all 
$\mid \varphi _j>\left( j=0,\cdots n-1\right) $. Next, from $\mu _{s,t}=%
\frac 1n\left( \delta _{st}-\left( N-n\right) p_s\right) <\Psi _s\mid \Psi
_t>\equiv 0$ this shows that: When $s\neq t,$ then $<\Psi _s\mid \Psi _t>$
must vanish, i.e. $\left\{ \mid \Psi _s>\right\} $ is a set of orthogonal
states; When $s=t$, it leads to $p_s\equiv \frac 1{N-n},$ hence $%
\sum\limits_{j=0}^{n-1}\mid \varphi _j><\varphi _j\mid +\sum\limits_s\rho
_s=I.$ To compare this with Eqs.(1) and (2), and since $\left\{ \mid \Psi
_s>\right\} $ is a set of orthogonal states, the index $s$ must run over $0$%
,1,$\cdots ,m-1,\left( m=N-n\right) .$ Sum up, we obtain a PB $S=\left\{
\mid \Psi _0>,\mid \Psi _1>,\cdots ,\mid \Psi _{m-1}>\right\} $ in the
orthogonal complementary subspace $H_T^{\bot }$. Further, this PB $S$ is an
UPB. In fact, if $\Omega $ is an arbitrary product state which is orthogonal
to all $\mid \Psi _i>\left( i=0,\cdots ,m-1\right) \in S,$ then it is in $%
H_T,$ because $H{\cal =}H_S\oplus H_T.$ But this is impossible, since $%
H_T=H_{ES}$ is an ES. This means that $S,$ in fact, is an UPB, i.e. $T$ is
an EEB and $H_{ES}$ is an EES. According to lemma 1, this $S$ is yet unique. 
$\square $

As for the problem how to decide the separability of a density matrix$,$ we
can use some criteria, e.g. the ways in [16,17], etc..

{\bf Corollary 1.} {\it An EB }$T=\left\{ \mid \varphi _0>,\cdots ,\mid
\varphi _{n-1}>\right\} (n\leq N-\sum\limits_{i=1}^M\left( d_i-1\right) -1)$%
{\it \ is an EEB if and only if the uniform mixture }$\stackrel{\_}{\rho }=%
\frac 1{N-n}\left\{ I-\sum\limits_{j=0}^{n-1}\mid \varphi _j><\varphi _j\mid
\right\} ${\it \ is separable.}

{\bf Proof.} According to the definition, the subspace $H_T$ spanned by $T$
is an EES if and only if $T$\ is an EEB, this means that corollary 1 holds. $%
{\bf \square }$

By this corollary, if $T$=$\left\{ \mid \varphi _0>,\cdots ,\mid \varphi
_{n-1}>\right\} $ is a EB and the uniform mixture $\stackrel{\_}{\rho }=%
\frac 1{N-n}\left\{ I-\sum\limits_{j=0}^{n-1}\mid \varphi _j><\varphi _j\mid
\right\} $ is separable, then it determines uniquely{\it \ }an UPB $%
S=\left\{ \mid \psi _0>,\cdots ,\mid \psi _{m-1}>\right\} $ and a CBUPB\ $%
B=S\cup T,$ where $\left\{ \mid \psi _0>,\cdots ,\mid \psi _{m-1}>\right\} $
is obtained from the above decomposition $\stackrel{\_}{\rho }=\sum_sp_s\rho
_s=\frac 1{N-n}\sum\limits_{i=0}^{n-1}\mid \psi _i><\psi _i\mid $ in the
proof of Theorem 1$.$

{\bf Corollary 2.} If a subspace $H_{ES}$ in $H$ is an EES, then for any
orthogonal basis $T=\left\{ \mid \varphi _0>,\cdots ,\mid \varphi
_{n-1}>\right\} $\ ($n\leq N-\sum\limits_{i=1}^M\left( d_i-1\right) -1)$ of $%
H_{ES}$ the uniform mixture $\stackrel{\backsim }{\rho }=\frac 1n%
\sum\limits_{j=0}^{n-1}\mid \varphi _j><\varphi _j\mid $ is a bound
entangled state[18].

{\bf Proof.} According to the above Theorem 1, now there is an UPB $%
S=\left\{ \mid \psi _0>,\cdots ,\mid \psi _{m-1}>\right\} \left(
m=N-n\right) $ and $\stackrel{\backsim }{\rho }=\frac 1{N-m}\left\{
I-\sum\limits_{i=0}^{m-1}\mid \psi _i><\psi _i\mid \right\} ,$ then this
corollary is just a known theorem (the Theorem 1$)$ in [1,3$].\;{\bf \square 
}$

We notice that a $\left( N-\sum\limits_{i=1}^{M-1}\left( d_i-1\right)
-1\right) $-dimensional EEB (or EES) is unextendible, i.e. there is no other
EEB (or EES) which contains this EEB (or EES) as a proper subset (or a
proper subspace). In fact, according to [1], in this case the dimensionality
of the UPB is minimum, so the dimensionality $\left(
N-\sum\limits_{i=1}^{M-1}\left( d_i-1\right) -1\right) $ of EEB (or EES) is
maximal.

Now we consider the problem what is the general structures of orthogonal
complete bases. Generally, in an orthogonal complete basis there may be some
entangled pure-states, especially, these entangled pure-states can form an
EB, even an EEB. All possible cases can be discussed as follows.

{\bf (I) In }$B${\bf \ there is no EB. }Now, a general orthogonal complete
basis of $H$ is in form as $B=S\cup R=\left\{ \mid \psi _0>,\cdots ,\mid
\psi _{m-1}>,\mid \Theta _0>,\cdots ,\mid \Theta _{N-m-1}>\right\} ,$ where $%
S=\left\{ \mid \psi _0>,\cdots ,\mid \psi _{m-1}>\right\} $ is an orthogonal
product basis, and $R=\left\{ \mid \Theta _0>,\cdots ,\mid \Theta
_{N-m-1}>\right\} $ is a set of orthogonal entangled pure-states, but $R$\
is not an EB, i.e. there may be a linear combination $\sum%
\limits_{i=0}^{N-m-1}c_i\mid \Theta _i>$ which is separable. This means that
if $m<N$, $S$ must be an extendible product basis. Here there are two
extreme cases: (i) $n=N$, then $B$ is a complete orthogonal product basis,
e.g. the common standard basis $\left\{ \mid j_{\alpha _1}>\otimes \cdots
\otimes \mid j_{\alpha _M}>\right\} $\ $(j_{\alpha _i}=0,1,\cdots ,d_i-1$ for%
$\;i=1,\cdots ,M)$ is such a basis$.$\ (ii) $n=0,$ then all pure-states in $%
B $ are entangled, e.g. the known Bell's basis in the bipartite qubit
systems is such a basis (notice that it is not an EB).

{\bf (II) In }$B${\bf \ there is an EB }$T{\cal =}\left\{ \mid \varphi
_0>,\cdots ,\mid \varphi _{n-1}>\right\} .$ Here, there are only two
possibilities: (i) The mixed-state $\stackrel{\_}{\rho }=\frac 1{N-n}\left\{
I-\sum\limits_{j=0}^{n-1}\mid \varphi _j><\varphi _j\mid \right\} $ is
separable, then according to the above theorem 1$,$ $T$ just is an EEB. This
means that there must be an UPB $S=$ $\left\{ \mid \psi _0>,\cdots ,\mid
\psi _{N-n-1}>\right\} ,$ and $B=S\cup T{\cal =}\left\{ \mid \psi _0>,\cdots
,\mid \psi _{N-n-1}>,\mid \varphi _0>,\cdots ,\mid \varphi _{n-1}>\right\} $
is an CBUPB. (ii) The mixed-state $\stackrel{\_}{\rho }=\frac 1{N-n}\left\{
I-\sum\limits_{j=0}^{n-1}\mid \varphi _j><\varphi _j\mid \right\} $ is
entangled. In order to discuss this case, we need to use the following lemma:

{\bf Lemma }${\bf 2}.$ {\it If }$T=\left\{ \mid \varphi _0>,\cdots ,\mid
\varphi _{n-1}>\right\} $ {\it is an EB and the mixed-state }$\stackrel{\_}{%
\rho }=\frac 1{N-n}\left\{ I-\sum\limits_{j=0}^{n-1}\mid \varphi _j><\varphi
_j\mid \right\} ${\it \ is entangled, \ then in the orthogonal complementary
space }$H_T^{\bot }${\it \ there is no }$\left( N-n\right) $-{\it %
dimensional orthogonal product basis.}

{\bf Proof. }Conversely{\bf , }suppose that in $H_T^{\bot }$ there is an
orthogonal product basis $S=\left\{ \mid \psi _0>,\cdots ,\mid \psi
_{N-n-1}>\right\} ,$ then $S$ must be an UPB, since $T$ is an EB. According
to the Theorem 1$,$ in this case $\stackrel{\_}{\rho }=\frac 1{N-n}\left\{
I-\sum\limits_{j=0}^{n-1}\mid \varphi _j><\varphi _j\mid \right\} $ must be
separable, but this contradicts to the supposition. $\square $

By using of Lemma 2, we know that in the case (ii) the form of an orthogonal
basis of $H_T^{\bot }$ must be as $\left\{ \mid \psi _0>,\cdots ,\mid \psi
_{r-1}>,\mid \phi _0>,\cdots ,\mid \phi _{s-1}>\right\} \left( 1\leq r\leq
N-n-2,0\leq s\leq N-n-1,r+s=N-n-1\right) $ , where $S=\left\{ \mid \psi
_0>,\cdots ,\mid \psi _{r-1}>\right\} $ is a PB$,$ and $T^{\prime }=\left\{
\mid \phi _0>,\cdots ,\mid \phi _{s-1}>\right\} $ is a set of orthogonal
entangled pure-states, but it is not an EB (i.e. some linear combination of $%
\mid \phi _0>,\cdots ,\mid \phi _{n-1}>$ may be a separable state). Now the
basis of $H$ is in form as $B=S\cup T^{\prime }\cup T$ =$\left\{ \mid \psi
_0>,\cdots ,\mid \psi _{r-1}>,\mid \phi _0>,\cdots ,\mid \phi _{s-1}>,\mid
\varphi _0>,\cdots ,\mid \varphi _{n-1}>\right\} .$ Here we notice that $%
T^{\prime }\cup T$ =$\left\{ \mid \phi _0>,\cdots ,\mid \phi _{s-1}>,\mid
\varphi _0>,\cdots ,\mid \varphi _{n-1}>\right\} $ and the basis $R$ in the
case (I) are very much alike, however in the former there is an EB in $B$.

Sum up, the forms of a general orthogonal complete bases $B$ can be
tabulated as follows: 
\[
\begin{array}{cccc}
\text{The first kind: In }B\text{ there is no EB} & - & - & 
\begin{array}{c}
\blacksquare \blacksquare \blacksquare \blacksquare \Vert ----\; \\ 
\text{(Special: }\blacksquare \blacksquare \blacksquare \blacksquare \Vert ,%
\text{ and }\Vert ----\text{)}
\end{array}
\\ 
&  &  &  \\ 
\text{The second kind: In }B\text{ there is an EB} & - & \text{(i) }%
\stackrel{\_}{\rho }\text{ is separable}- & 
\begin{array}{c}
\Box \Box \Box \Box \Vert \equiv \equiv \equiv \equiv \\ 
\text{(it is a CBUPB by a proper choice of }S\text{)}
\end{array}
\\ 
& - & \;\text{(ii) }\stackrel{\_}{\rho }\text{ is entangled}- & \blacksquare
\blacksquare \blacksquare \blacksquare \Vert ----====
\end{array}
\]
where $\blacksquare \blacksquare \blacksquare \blacksquare $ denotes a PB; $%
----$ denotes the basis consisting of entangled pure-states, but it is not
an EB; $\Box \Box \Box \Box $ denotes an UPB; $====$ denotes an EB, but it
is not an EEB; $\equiv \equiv \equiv \equiv $ denotes an EEB. The symbol $%
\Vert $ denotes the partition between two parts PB and EB.

{\bf Applications and examples in quantum mechanics and quantum information}

{\bf 1.} {\bf A }CBUPB{\bf \ constructed from the GenTiles1. }About the
requirements and applications of the PBs and UPBs in the quantum
information, see $[2,3].$ A known UPB is the {\bf GenTiles} given by
DiVincenzo, Terhal et al.[2,3]. By using of the symbols as in [2,3], in $H%
{\cal =}H_n\otimes H_n$ $(n$\ is even and $n\geq 4)$\ the {\bf GenTiles1} $%
\left\{ \mid {\bf V}_{mk}>,\mid {\bf H}_{mk}>,\mid {\bf F}>\right\} $ is
defined by (here we have taken their normalization).

\begin{eqnarray}
&\mid &{\bf V}_{mk}>=\frac 2n\mid k>\otimes \mid \omega _{m,k+1}>=\frac 2n%
\mid k>\otimes \sum_{j=0}^{\frac n2-1}\omega ^{jm}\mid j+k+1\text{ mod }n>,%
\text{ }\omega =e^{\frac{i4\pi }n}  \nonumber \\
&\mid &{\bf H}_{mk}>=\frac 2n\mid \omega _{m,k}>\otimes \mid k>,\;\mid {\bf F%
}>=\frac 1{n^2}\sum_{i=0}^{n-1}\sum_{j=0}^{n-1}\mid i>\otimes \mid j>
\end{eqnarray}
where $\;m=1,\cdots ,\frac n2-1,\;k=0,\cdots ,n-1.$If we take $\mid {\bf F}%
_r>=\mid r>\otimes \mid r>,\;(r=0,1,\cdots ,2\left( n-1\right) ),$\ then $%
\left\{ \mid {\bf V}_{mk}>,\mid {\bf H}_{mk}>,\mid {\bf F}>,\mid {\bf F}%
_r>\right\} $\ (the total of states is $n^2$) is a linearly independent
group of normalized states. Now, we define the normalized states $\mid
\varphi _r>(r=0,\cdots ,2\left( n-1\right) )$\ by induction as follows (i.e.
the Schmidt's orthogonalization) 
\begin{eqnarray}
&\mid &\varphi _0>=\eta _0\left\{ \mid {\bf F}_0>-\sum_{m=1}^{\frac n2%
-1}\sum_{k=0}^{n-1}\left( <{\bf V}_{mk}\mid {\bf F}_0>\mid {\bf V}_{mk}>+<%
{\bf H}_{mk}\mid {\bf F}_0>\mid {\bf H}_{mk}>\right) -<{\bf F}\mid {\bf F}%
_0>\mid {\bf F}>\right\}  \nonumber \\
&=&\eta _0\left( \mid 0>\otimes \mid 0>-\frac 2n{\bf H}_{00}-\frac 1{n^2}%
\mid {\bf F}>\right) ,\eta _0=\frac{4n^8}{4\left( n^4-2n^2-1\right)
^2+\left( n-2\right) ^2\left( 4n^2+1\right) ^2+4n^2\left( n-1\right) ^2+n^2}
\\
\varphi _r &=&\eta _r\left\{ \mid {\bf F}_r>-\sum\limits_{m=1}^{\frac n2%
-1}\sum\limits_{k=0}^{n-1}\left( <{\bf V}_{mk}\mid {\bf F}_r>\mid {\bf V}%
_{mk}>+<{\bf H}_{mk}\mid {\bf F}_r>\mid {\bf H}_{mk}>\right) -<{\bf F}\mid 
{\bf F}_r>\mid {\bf F}>-\sum\limits_{s=0}^{r-1}<\varphi _s\mid {\bf F}%
_r>\mid \varphi _s>\right\}  \nonumber \\
&=&\eta _r\left\{ \mid {\bf F}_r>-\frac 2n{\bf H}_{rr}-\left(
1+\sum\limits_{s=0}^{r-1}\eta _s\right) \frac 1{n^2}\mid {\bf F}>\right\}
,\;r=1,\cdots ,2\left( n-1\right)  \nonumber
\end{eqnarray}
where $\eta _r(r=0,\cdots ,2\left( n-1\right) )$\ are normalization factors,
which are also determined by induction. Therefore the set $\left\{ \mid {\bf %
V}_{mk}>,\mid {\bf H}_{mk}>,\mid {\bf F}>,\mid \varphi _r>\right\} $\ $%
\left( \;m=1,\cdots ,\frac n2-1,\;k=0,\cdots ,n-1,\;r=0,\cdots ,2n-1\right) $%
\ forms an orthogonal complete basis of $H=H_n\otimes H_n.$\ Since the {\bf %
GenTiles1} $\left\{ \mid {\bf V}_{mk}>,\mid {\bf H}_{mk}>,\mid {\bf F}%
>\right\} $\ is an UPB[2,3] and $\left\{ \mid \varphi _r>\right\} $\ is in $%
H_{{\bf GenTiles1}}^{\bot }$\ $(H_{{\bf GenTiles1}}$\ is the subspace
spanned by the {\bf GenTiles1}), every $\mid \varphi _r>$\ is entangled, the
set $\left\{ \mid \varphi _r>\right\} $\ is an EEB, and $\left\{ \mid {\bf %
V\;\;}_{mk}>,\mid {\bf H}_{mk}>,\mid {\bf F\;}>,\mid \varphi _r>\right\} $\
is a CBUPB\ . In addition, according to the Theorem 1 and Corollary 2, \ $%
\stackrel{\_}{\rho }=\frac 1{\left( n-1\right) ^2}\left\{
I-\sum\limits_{r=0}^{2\left( n-1\right) }\mid \varphi _r><\varphi _r\mid
\right\} $\ is separable, and $\stackrel{\backsim }{\rho }=\frac 1{2n-1}%
\sum\limits_{r=0}^{2\left( n-1\right) }\mid \varphi _r><\varphi _r\mid $\ is
a bound entangled state.

This example clearly shows the use of the {\bf GenTiles1 }and the theorem 1.
A similar result can be obtained from the {\bf GenTiles2}[2,3].

{\bf 2. Some bound entangled states. }In the quantum information, one need
to study the problem what the entanglement can be distilled, because this
relates the quantum communication problems. A bound entangled state is such
a entangled state that no entanglement can be distilled[18]. If in the
system there is an EEB $T=\left\{ \mid \varphi _0>,\cdots ,\mid \varphi
_{n-1}>\right\} ,$ then according to the above Corollary 2, the uniform
mixture $\stackrel{\backsim }{\rho }=\frac 1n\sum\limits_{j=0}^{n-1}\mid
\varphi _j><\varphi _j\mid $ is entangled and is a bound entangled state,
this is a standard case as in [1,3]. However, we find that there may be yet
other bound entangled states with similar forms as follows. For instance,
suppose that in the system there is a EB $T=\left\{ \mid \varphi _0>,\cdots
,\mid \varphi _{n-1}>\right\} ,$ but the mixed-state $\stackrel{\_}{\rho }=%
\frac 1{N-n}\left\{ I-\sum\limits_{j=0}^{n-1}\mid \varphi _j><\varphi _j\mid
\right\} $ is entangled, then according to the above classification tab,
there are a PB $S=\left\{ \mid \psi _0>,\cdots ,\mid \psi _{r-1}>\right\} $
and a $T^{\prime }=\left\{ \mid \phi _0>,\cdots ,\mid \phi _{s-1}>\right\}
\left( r+s+n=N\right) ,$ which is a set of orthogonal entangled pure-states
but not an EB, they form together an orthogonal complete basis $B=S\cup
T^{\prime }\cup T$. Let 
\begin{equation}
\widetilde{\rho ^{\prime }}=\frac 1{s+n}\left\{ \sum_{i=0}^{s-1}\mid \phi
_k><\phi _k\mid +\sum_{j=0}^{n-1}\mid \varphi _j><\varphi _j\mid \right\} =%
\frac 1{N-r}\left\{ I-\sum_{k=0}^{r-1}\mid \psi _k><\psi _k\mid \right\}
\end{equation}
Since $\widetilde{\rho ^{\prime }}$ contains the component part constructed
by the states in $T^{\prime }$, we cannot affirm that $\widetilde{\rho
^{\prime }}$ must be a entangle state for any $T^{\prime }$. However in many
cases, by a proper choice of $T^{\prime }$, $\widetilde{\rho ^{\prime }}$
always can become entangled. Notice that the related argument in the proof
about $\stackrel{\backsim }{\rho }$ to be a bound entangled state (see the
proof of Theorem 1 in [1,3]), obviously, is still completely effective for $%
\widetilde{\rho ^{\prime }}$ , i.e. in present case also no entanglement can
be distilled. Therefore $\widetilde{\rho ^{\prime }}$ is a bound entangled
state. This example shows the multiformity of bound entangled states with
the uniform mixture forms.

At last, we consider the problem of transformations. Since at present we
only discuss the orthogonal bases, we cannot use the general local
operations and classical communication[13]. However, we can use the local
unitary transformations.

{\bf Theorem 2.} The structure form of a general orthogonal complete basis $%
B=S\cup T{\cal =}\left\{ \mid \psi _0>,\cdots ,\mid \psi _{n-1}>,\mid
\varphi _0>,\cdots ,\mid \varphi _{k-1}>\right\} \left( n+k=N\right) $ is
invariant under a local operation as 
\begin{eqnarray}
G &:&B\longrightarrow B^{\prime }=S^{\prime }\cup T^{\prime }{\cal =}\left\{
\mid \psi _0^{\prime }>,\cdots ,\mid \psi _{n-1}^{\prime }>,\mid \varphi
_0^{\prime }>,\cdots ,\mid \varphi _{k-1}^{\prime }>\right\}  \nonumber \\
&\mid &\psi _i^{\prime }>=u\left( d_1\right) \otimes u\left( d_2\right)
\otimes \cdots \otimes u\left( d_M\right) \left( \mid \psi _i>\right) ,\mid
\varphi _j^{\prime }>=u\left( d_1\right) \otimes u\left( d_2\right) \otimes
\cdots \otimes u\left( d_M\right) \left( \mid \varphi _j>\right)
\end{eqnarray}
where $u\left( d_i\right) (\left( i=1,\cdots ,M\right) $ are arbitrary $%
d_i\times d_i$ unitary matrixes.

{\bf Proof.} Obviously, an interior product of any two states is invariant
under this $G$, then an orthogonal complete basis will be changed into
another orthogonal complete basis under $G$, i.e. $G$ indeed is a
transformation of general orthogonal complete bases. Next, for a pure-state $%
\mid \psi >,$ the pure-state $\mid \psi ^{\prime }>=u\left( d_1\right)
\otimes u\left( d_2\right) \otimes \cdots \otimes u\left( d_M\right) \left(
\mid \psi >\right) $ is a product state, if and only if $\mid \psi >$ is a
product state itself. This means that if in $B$ there is no EB, then in $%
B^{\prime }=G\left( B\right) $ there is yet no EB. In addition, when in $B$
there is an\ EB $T,$ the corresponds uniform mixture is $\stackrel{\_}{\rho }%
,$ then the uniform mixture $\stackrel{\_}{\rho }^{\prime }=u\left(
d_1\right) \otimes u\left( d_2\right) \otimes \cdots \otimes u\left(
d_M\right) \stackrel{\_}{\rho }u^{-1}\left( d_M\right) \otimes u^{-1}\left(
d_{M-1}\right) \otimes \cdots \otimes u^{-1}\left( d_1\right) $ is
separable, if and only if $\stackrel{\_}{\rho }$ is separable itself.
According to the above classification tab and these discussions, it has been
proved that the structure form of a general orthogonal complete basis $B$ is
invariant under $G.$ ${\bf \Box }$

This theorem can be used to create various orthogonal complete bases (in the
same type) from given one.

\end{document}